\documentclass{ifacconf}

\usepackage{amssymb}
\usepackage{amsmath}
\usepackage{amsfonts}                                
\usepackage[latin1] {inputenc}
\usepackage{enumerate}
\usepackage{mathrsfs}
\usepackage{tabulary}
\usepackage{cite}
\usepackage{psfrag}
\usepackage{cite}
\usepackage{graphicx}          
\usepackage{color}      
\usepackage{epsfig} 
\usepackage{epstopdf}
\usepackage{natbib}        
\usepackage{times} 
\def\qedp{\hspace*{\fill}~{\tiny $\blacksquare$}}
\def\qed{\relax\ifmmode\hskip2em \Box\else\unskip\nobreak\hskip1em $\Box$\fi}

\newtheorem{theorem}{Theorem}
\newtheorem{itlemma}{Lemma}
\newtheorem{itdefinition}{Definition}
\newtheorem{itproposition}{Proposition}
\newtheorem{itresult}{Result}
\newtheorem{itremark}{Remark}
\newtheorem{itassumption}{Assumption}
\newtheorem{itcorollary}{Corollary}
\newtheorem{itexample}{Example}

\newenvironment{remark}{\begin{itremark}\rm}{\end{itremark}}
\newenvironment{assumption}{\begin{itassumption}\rm}{\end{itassumption}}
\newenvironment{lemma}{\begin{itlemma}\rm}{\end{itlemma}}

\begin{document}
\begin{frontmatter}

\title{Networked Systems under Denial-of-Service: Co-located vs. Remote Control Architectures} 

\author[First]{Shuai Feng} 
\author[First]{Pietro Tesi} 

\address[First]{ENTEG, Faculty of Science and Engineering, University of Groningen, 9747 AG Groningen, 
The Netherlands (e-mail: \{s.feng,p.tesi\}@rug.nl).}

\begin{abstract}
In this paper, we study networked systems in the presence 
of Denial-of-Service (DoS) attacks, namely attacks that prevent transmissions 
over the communication network. Previous studies have shown that co-located architectures 
(control unit co-located with the actuators and networked sensor channel)
can ensure a high level of robustness against DoS. 
However, co-location requires a wired or dedicated actuator channel, 
which could not meet flexibility and cost requirements. 
In this paper we consider a control architecture that approximates co-location while enabling remote implementation
(networked sensor and actuator channels).
We analyze closed-loop stability and quantify the robustness ``gap"
between this architecture and the co-located one.
\end{abstract}


\end{frontmatter}

\section{introduction}	

The field of cyber-physical systems is becoming more and more important 
in control engineering and computer science due to its broad spectrum of applications.
Especially for safety-critical applications, there is a 
need for analysis, synthesis and design tools that can
guarantee secure and reliable operations despite the presence of malicious attacks \citep{SpecialIssue17}.

Security of cyber-physical systems involves several research areas, including 
anomaly detection, verification, and estimation and control in the presence of attacks. Moreover, 
the problem varies depending on the type of attack one is considering.
One usually classifies cyber attacks as either deception attacks, examples being 
zero-dynamics and bias injection attacks or Denial-of-Service (DoS) attacks. 
The former affect the data trustworthiness via the manipulation of packets 
transmitted through the network \citep{Fawzi,Teixeira2015135}. 
DoS attacks are instead intended to affect the timeliness 
of the information exchange, \emph{i.e.}, to induce packet losses
\citep{xu2005feasibility,sastry}.
In particular, it is known that DoS attacks are relatively easy to launch 
\citep{xu2005feasibility}, even without knowledge of the targeted system. 

This paper deals with DoS attacks. 
We consider a control system in which plant-controller
communication is networked. An attacker can induce closed-loop instability 
by denying the communication on sensor and actuator channels. 
The problem of interest is to design control systems that can
tolerate DoS as much as possible.
Networked control in the presence of packet losses has been widely investigated \citep{Naghshtabrizi}.
However, it is known that communication failures induced by DoS can exhibit a temporal profile 
quite different from the one induced by genuine packet losses; 
in particular, communication failures induced by DoS need not follow a given class of probability distributions \citep{sastry}.
This raises new theoretical challenges from the perspective of analysis as well as control design.

The literature on networked control under DoS is large and quite diversified.
In \citep{basar,sastry}, the authors address 
the problem of finding optimal control and DoS attack strategies 
assuming a maximum number of jamming actions over a prescribed control horizon. A similar 
problem is considered in \citep{Ugrinovskii}, where the authors 
study zero-sum games between controllers and  strategic jammers. 
In \citep{zhang2016optimal}, the authors investigate DoS from the attacker's viewpoint, 
and characterize optimal DoS attack scheduling strategies.
\cite{martinez2} consider periodic DoS attacks. 
The objective is to identify salient features of the DoS signal such as maximum 
\emph{on/off} cycle in order to de-synchronize
the transmission times from the occurrence of DoS. 
In \citep{CDP:PT:IFAC14,de2015input}, a framework is introduced where
no assumption is made regarding the ``structure" of the DoS attack.  
Instead, a model is considered that constrains
DoS only in terms of its \emph{frequency} and \emph{duration}. 
The contribution is an explicit characterization of 
DoS frequency and duration 
for which stability can be preserved
through {state-feedback control}. 
Extensions have been considered 
dealing with co-located robust controller design \citep{Feng201742},
nonlinear \citep{CDP:PT:CDC14} and distributed \citep{senejohnny} systems,
as well as with systems where jamming attacks and genuine packet losses coexist \citep{7574308}.

In a recent paper \citep{7526102}, we have investigated networked systems under DoS attacks
from the perspective of designing ``maximally robust" controllers. There, it is shown that 
co-located architectures (control unit co-located with the actuators and networked sensor channel)
can guarantee the highest possible level of robustness against DoS,
in the sense that they can guarantee closed-loop stability for all the
DoS signals with frequency and duration below a certain critical threshold 
beyond which stability can be lost irrespective of the adopted control system.
Unfortunately, co-location requires wired or dedicated actuator channels, 
which could not meet flexibility and cost requirements. 

In this paper, we consider a control system that  
approximates co-location while enabling remote implementation
(networked sensor and actuator channels).
The proposed architecture relies on packet-based control and actuator buffering.
The basic idea is to transmit data packets containing plant input predictions whenever 
DoS is absent, and use the data stored in the actuator buffer during the periods of DoS.
This idea has already proved effective to compensate for network delay 
\citep{chaillet2008delay,Bemporad} as well as packet losses \citep{quevedo2011input2,quevedo2011input}. 
In this paper, however, the peculiarity of the problem leads to a different analysis and design.
We follow a ``worst-case" type of analysis, accounting for the situation 
where the network undergoes periods of DoS much larger than the buffer capacity.

The paper contribution is twofold. First, we provide conditions 
on the prediction horizon under which closed-loop stability is preserved.
The analysis is general enough to account for: (i) process disturbances; (ii)
measurement and network noise; (iii) non-zero computation times. 
As a second contribution, we explicitly quantify the robustness ``gap" between this 
control architecture and the co-located one,
showing that the ideal bound obtained with co-location is recovered
as the prediction horizon increases.   

\subsection{Notation}
We denote by $\mathbb R$ the set of reals. For any 
$\alpha \in \mathbb R$, we denote $\mathbb R_{\geq \alpha}:=\{x \in \mathbb R: x \geq \alpha\}$.
We let $\mathbb N_0$ denote the set of nonnegative integers, 
$\mathbb N_0 := \{0,1,\ldots\}$. For any $\alpha \in \mathbb N_0$, we denote
$\mathbb N_\alpha := \{\alpha,\alpha+1,\ldots\}$. The prime denotes transpose.
Given a vector $v \in \mathbb R^n$, $\|v\|$ is its Euclidean norm. 
Given a matrix $M$, $\|M\|$ is its spectral norm. 
Given a measurable time function 
$f:  \mathbb R_{\geq 0} \mapsto  \mathbb R^n$ 
and a time interval $[0,t)$
we denote the $\mathcal L_\infty$ norm of $f(\cdot)$ on $[0,t)$ 
by $\|f_t\|_{\infty} := \textrm{sup}_{s \in [0,t]} \|f(s)\|$. 
Given a measurable time function 
$f:  \mathbb R_{\geq 0} \mapsto  \mathbb R^n$ we say 
that $f$ is bounded if its $\mathcal L_\infty$ norm is finite.

\section{Framework and paper outline}\label{sec.problem}

\subsection{Process dynamics and network}

The process to be controlled 
is given by 
{\setlength\arraycolsep{2pt} 
	\begin{eqnarray} \label{system}
	\left\{ \begin{array}{rl}
	\dot x(t) & = A x(t) + B u(t) + d(t) \\
	y(t) & = x(t) + n(t) \\
	x(0) &= x_0
	\end{array} \right.
	\end{eqnarray}}%
where $t \in \mathbb R_{\geq 0}$; $x \in \mathbb R^{n}$ is the state,
$u \in \mathbb R^{m}$ is the control input and
$y \in \mathbb R^{n}$ is measurement vector; $A$ and $B$ are 
matrices of appropriate size with $(A,B)$ stabilizable; $d \in \mathbb R^{n}$ and $n \in \mathbb R^{n}$ 
are unknown (bounded) disturbance and noise
signals, respectively. 

We assume that sensor and actuator channels are networked
and subject to Denial-of-Service (DoS) status.
The former implies that measurements and control commands are sent 
only at discrete time instants. Let 
$\{t_k\}_{k \in \mathbb N_0} = \{t_0,t_1,\ldots\}$
denote the sequence of transmission attempts. 
Throughout the paper, we assume for simplicity 
that the transmission attempts are carried out periodically 
with period $\Delta$, \emph{i.e.},
\begin{eqnarray} \label{eq:transmission_attempts}
t_{k+1} -t_{k}=\Delta, \quad k \in \mathbb N_0
\end{eqnarray}
with $t_0=0$ by convention. 
We refer to DoS as the phenomenon for which 
some transmission attempts may fail. 
	
We shall denote by $\{z_m\}_{m \in \mathbb N_0} = \{z_0,z_1,\ldots\}$, 
$z_0 \geq t_0$, the sequence of time instants at which samples of $y$ are successfully transmitted. 

\subsection{Control objective} 

The objective is to design a controller $\mathcal K$, possibly dynamic, in such a way 
that the closed-loop stability is maintained despite the occurrence of DoS in measurement 
and control channels. In this paper, by closed-loop stability we mean that all the signals in 
the closed-loop system remain bounded for any initial condition $x_0$ and bounded noise 
and disturbance signals, and converge to zero in the event that noise and disturbance signals converge to zero.

\subsection{Denial-of-Service: Assumptions} \label{sub:DoS}

Clearly, the problem in question does not have a solution 
if the DoS amount is allowed to be arbitrary.
Following \citep{de2015input}, we consider a general DoS model
that constrains the attacker action in time 
by only posing limitations on the frequency of DoS attacks and their duration.
Let 
$\{h_n\}_{n \in \mathbb N_0}$, $h_0 \geq 0$, denote the sequence 
of DoS \emph{off/on} transitions, \emph{i.e.},
the time instants at which DoS changes from zero (transmissions are possible) to one 
(transmissions are not possible).
Hence, $H_n :=\{h_n\} \cup [h_n,h_n+\tau_n[$
represents the $n$-th DoS interval, of length 
$\tau_n \in \mathbb R_{\geq 0}$,
over which the network is in DoS status. If $\tau_n=0$, then
$H_n$ takes the form of a single pulse at $h_n$.  
Given $\tau,t \in \mathbb R_{\geq0}$ with $t\geq\tau$, 
let $n(\tau,t)$
denote the number of DoS \emph{off/on} transitions
over $[\tau,t[$, and let 
\begin{eqnarray}  \label{DoS_intervals_union}
\Xi(\tau,t) := \bigcup_{n \in \mathbb N_0} H_n  \, \bigcap  \, [\tau,t] 
\end{eqnarray}
denote the 
subset of $[\tau,t]$ where the network is in DoS status. 

We make the following assumptions.

\begin{assumption}
	(\emph{DoS frequency}). 
	There exist constants 
	$\eta \in \mathbb R_{\geq 0}$ and 
	$\tau_D \in \mathbb R_{> 0}$ such that
\begin{eqnarray} \label{ass:DoS_slow_frequency} 
n(\tau,t)  \, \leq \,  \eta + \frac{t-\tau}{\tau_D}
\end{eqnarray}
	for all  $\tau,t \in \mathbb R_{\geq0}$ with $t\geq\tau$.
	\qedp
\end{assumption}

\begin{assumption} 
	(\emph{DoS duration}). 
	There exist constants $\kappa \in \mathbb R_{\geq 0}$ and $T  \in \mathbb R_{>1}$ such that
\begin{eqnarray} \label{ass:DoS_slow_duration}
|\Xi(\tau,t)|  \, \leq \,  \kappa + \frac{t-\tau}{T}
\end{eqnarray}
	for all  $\tau,t \in \mathbb R_{\geq0}$ with $t\geq\tau$. 
	\qedp
\end{assumption}

\begin{remark}
Assumptions 1 and 2 do only constrain a given DoS signal in terms of its \emph{average} frequency and duration.
In practice, 
$\tau_D$ can be defined as the average dwell-time between 
consecutive DoS off/on transitions, while $\eta$ is the chattering bound.
Assumption 2 expresses a similar 
requirement with respect to the duration of DoS. 
It expresses the property that, on the average,
the total duration over which communication is 
interrupted does not exceed a certain \emph{fraction} of time,
as specified by $1/T$.
Like $\eta$, the constant $\kappa$ plays the role
of a regularization term. It is needed because
during a DoS interval, one has $|\Xi(h_n,h_n+\tau_n)| = \tau_n >  \tau_n /T$.
Thus $\kappa$ serves to make (\ref{ass:DoS_slow_duration}) consistent. 
Conditions $\tau_D>0$ and $T>1$ imply that DoS cannot occur at an infinitely
fast rate or be always active. \qedp
\end{remark}
 
\subsection{Previous work and paper outline: 
co-located and remote control architectures}

In \cite{7526102}, we have investigated the problem 
of designing control architectures that achieve \emph{maximal robustness} against DoS.
We briefly summarize the main result of that paper.
Consider the control architecture depicted in Figure 1. The control system is 
\emph{co-located} with the process actuators and is equipped  
with prediction capabilities so as to reconstruct the missing measurements during DoS periods.
Let $\delta:={\Delta}/{b} $
be the sampling rate of the control system,
where $b \in \mathbb N_{1}$. 
Notice that choosing $\delta=\Delta/b$ allows to differentiate 
between controller sampling rate and transmission rate,
maintaining $\Delta$ possibly large. 
Finally, let $A_\delta :=e^{A\delta}$ and $B_\delta:=\int_{0}^{\delta} e^{A \tau} B d\tau$.
The predictor equations are given by
{\setlength\arraycolsep{2pt} 
\begin{eqnarray}  \label{co-located predictor}
\arraycolsep=1.4pt\def\arraystretch{1.7}
\left\{ \begin{array}{l}
\xi((q+1)\delta) = A_\delta \alpha(q\delta)+B_\delta u(q\delta) \\ 
\alpha(q\delta) = \left\{ \begin{array}{rl}
y(q\delta), & \quad \textrm{if } q\delta=z_m \\
\xi(q\delta), & \quad \textrm{otherwise} 
\end{array} \right. \\
\xi (0) = 0 
\end{array} \right.
\end{eqnarray}}%
where $q \in \mathbb N_0$, and the control action is given by
\begin{eqnarray} \label{co-located feedback}
u(t) =K\alpha(q\delta), \quad t \in [q\delta,(q+1)\delta[ 
\end{eqnarray}

Here $K$ is a state-feedback matrix such that all the eigenvalues of $\Phi=A+BK$ 
have negative real part. The control system consists of a predictor and a state-feedback matrix. The
former emulates the process dynamics during the DoS period, and is ``reset" whenever 
new measurements become available.

\begin{theorem}[\cite{7526102}]
	Consider a process as in (\ref{system}) under a co-located control system as in 
	(\ref{co-located predictor}) and (\ref{co-located feedback}). 
	Given any positive definite symmetric matrix $M$,
	let $P$ denote the solution of the Lyapunov equation 
	$\Phi' P + P \, \Phi + M = 0$. 
	Let the controller sampling rate $\delta$ 
	be such that
	\begin{eqnarray} \label{eq:delta_1}
	\delta  \,  \leq \, 
	\frac{1}{\mu_A} 
	\log \left[ \left( \frac{\sigma}{1+\sigma} \right) \frac{ \mu_A }{\max \{\|\Phi\|,1\}} +1 \right]
	\end{eqnarray}
	when $\mu_A > 0$, and
	\begin{eqnarray}  \label{eq:delta_2}
	\delta  \,  \leq \, 
	\left( \frac{\sigma}{1+\sigma} \right) \frac{ 1}{\max\{\|\Phi\|,1\}}
	\end{eqnarray}
	when $\mu_A \leq 0$, where $\mu_A$ is the logarithmic norm of $A$ and 
	$\sigma$ is a positive constant satisfying
	$
	\gamma_1  - \sigma \gamma_2 >0
	$,
	where $\gamma_1$ is equal to the smallest
	eigenvalue of $M$ and $\gamma_2:=\|2PBK\|$. 
	Then, the closed-loop system is stable
	for any DoS sequence 
	satisfying Assumptions 1 and 2 with
	arbitrary $\eta$ and $\kappa$, and with $\tau_D$ and $T$ such that
	\begin{eqnarray} \label{co-located controller bound of DoS}
	\frac{1}{T} + \frac{\Delta}{\tau_D} \, < \,  1
	\end{eqnarray} 
	\qedp
\end{theorem}

\begin{figure}[t]
	\begin{center}
		\includegraphics[width=0.40 \textwidth]{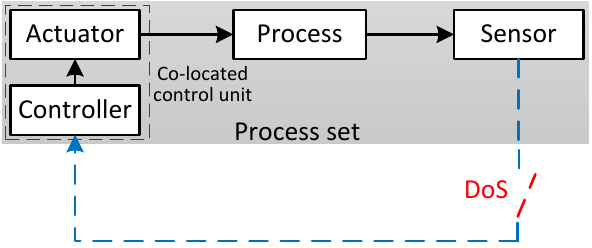} \\
		\linespread{1}\caption{\small Co-located control architecture.} 
	\end{center}
\end{figure}

The bound (\ref{co-located controller bound of DoS}) is the {best} bound that one can achieve
for the class of DoS signals satisfying Assumptions 1 and 2
in the sense that
at or above the threshold ``1" (when $1/T + \Delta/\tau_D \geq 1$) 
one can preserve stability only for \emph{some} DoS signals, not for \emph{all}.
In fact, as soon as we reach the threshold ``1", 
$\tau_D$ and $T$ can give rise to DoS signals that disrupt all the transmission attempts.
Consider, for example, the DoS signal
given by $(h_n,\tau_n)=(n \Delta,0)$. This signal 
yields $1/T + \Delta/\tau_D=1$ and satisfies 
Assumption 1 and 2 with $\eta=1$, $\kappa=0$,
$T=\infty$ and $\tau_D = \Delta$.
By construction, this DoS signal disrupts all the transmission attempts
since it is synchronized with the transmission times.
Note that condition $1/T + \Delta/\tau_D<1$ requires
$\tau_D > \Delta$. This means that, on the average, 
DoS cannot occur at the same rate as (or faster than) $\Delta$. 

Despite its robustness, the control architecture in (\ref{co-located predictor})-(\ref{co-located feedback})
has a number of practical shortcomings: \emph{co-location} requires a wired or dedicated control channel, 
which could not meet flexibility and cost requirements. In order to mitigate these shortcomings, we consider an 
architecture that allows for \emph{remote} control. As shown in Figure 2, we consider 
a ``packetized-and-buffered" architecture in which the controller transmits a sequence of control values 
containing process input predictions. Whenever communication is available, these values 
are stored in a buffer and used during the DoS periods. In the remainder of this paper,
we formally investigate stability and robustness properties of this architecture.
We show that under (\ref{co-located controller bound of DoS})
closed-loop stability is preserved as long as the prediction horizon $h \delta$ satisfies
\begin{eqnarray} \label{eq:stability_condition}
h \delta > \frac{\omega_2}{\omega_1+\omega_2} (Q+\Delta) 
\end{eqnarray}
where $h$ is the buffer size, $\delta$ is the actuator sampling interval, 
$\omega_1$ and $\omega_2$ are process-dependent constants, 
and $Q$ is a DoS-dependent constant. The above stability condition can
can be equivalently rewritten as
\begin{eqnarray} \label{eq:stability_condition_equivalent}
\frac{1}{T}+\frac{\Delta}{\tau_D} < 1- \frac{\omega_2(\kappa+\eta\Delta)}{(\omega_1+\omega_2)h\delta-\omega_2\Delta}
\end{eqnarray}

This inequality explicitly quantifies the gap between  
co-located and remote architectures, showing that the latter approaches
the ideal bound  as $h \rightarrow \infty$.

\begin{figure}[t]
	\begin{center}
		\includegraphics[width=0.40 \textwidth]{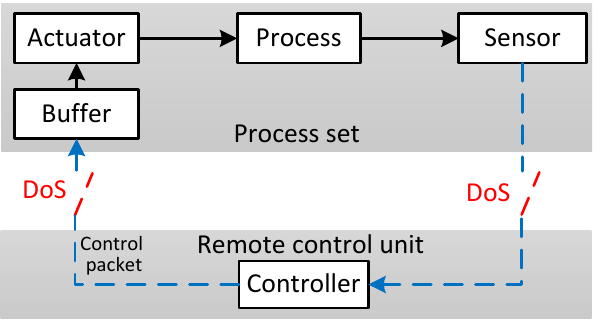} \\
		\linespread{1}\caption{\small Remote control architecture.} 
	\end{center}
\end{figure}

\section{Remote control architecture}

The actuator implements a \emph{sample-and-hold} control strategy
with sampling period $\delta := \Delta/b$, $b\in \mathbb{N}_{1}$. 
The control unit operates at the time instants $z_m$ at which 
process measurements are available, returning $h \in \mathbb{N}_{1}$ controls
\begin{eqnarray} \label{control samples}
u_p(z_m) = K \alpha_p(z_m), \quad p= 0,\ldots,h-1 
\end{eqnarray}

where the variable $\alpha_p(z_m)$ defines the prediction of $x(z_m+p\delta)$ 
at time $z_m$.
In particular, $\alpha_p(z_m)$ satisfies
{\setlength\arraycolsep{2pt} 
	\begin{eqnarray}  \label{eq:dt_predictor}
	\arraycolsep=1.4pt\def\arraystretch{1.5}
	&& \alpha_0(z_m) = y(z_m), \\
	&& \alpha_{p+1}(z_m) = A_\delta \alpha_p(z_m)+B_\delta u_p(z_m) \
\end{eqnarray}}%

where $p \in \{0,\ldots,h-2\}$ and where $A_\delta$ and $B_\delta$ are as in (\ref{co-located predictor}).
In practice, the control unit implements a sampled-data version of the process dynamics,
which are ``reset" whenever a new process measurement becomes available.

Let $I_{p}(z_m) :=[z_m+p\delta,\min\{z_m+(p+1)\delta,z_{m+1}\}[$, where
$p \in \{0,\ldots,h-2\}$, and $I_{h-1}(z_m) :=[z_m+(h-1)\delta,z_{m+1}[$.
The control input applied to the process is
\begin{eqnarray} {\label{control input}}
\arraycolsep=1.4pt\def\arraystretch{1.3}
u(t) = \left\{ \begin{array}{lll} 0, & \,\,\, t  \in [0,z_0[  \smallskip   \\
u_p(z_{m}), & \,\,\,  t \in I_{p}(z_{m}), \,\,\, p =0,\ldots,h-2 \smallskip   \\
u_{h-1}(z_m), &  \,\,\, t \in I_{h-1}(z_{m}) 
\end{array} \right. \nonumber \\
\end{eqnarray}
In words, the control action is kept to zero until the first process measurement is received.
Thereafter, as shown in Figure 3, over each interval $[z_m,z_{m+1}[$ the actuator applies the values in the buffer
until the buffer is empty and the last value in the buffer $u(z_m+(h-1)\delta)$ is kept until $z_{m+1}$.
On the other hand, if $z_{m+1} < z_m+h\delta$ then all the samples in the buffer are discarded
and a new sequence of controls is stored. This renders (\ref{control input}) a ``receding horizon" control policy.

Notice that (\ref{control input}) assumes that there is no noise in the actuator channel.
This is because, if the actuator receives $u_p + n_u$, where $n_u$ is a noise, 
the contribution of $n_u$ can be absorbed in the process disturbance $d$. 
Also notice that (\ref{control input}) assumes that the time needed to compute all the 
samples in (\ref{control samples}) is zero. This assumption can be relaxed as discussed 
in Section 4.

\section{Main result}

We first present a lemma which is key to our developments. 
Then, we analyze the closed-loop behavior within and outside the 
prediction horizon and provide stability conditions. 
 
\subsection{Key lemma}

The lemma relates DoS parameters and 
time elapsing between successful transmissions.

\vspace{0.2cm}

\begin{lemma}[\cite{7526102}]
Consider a transmission policy as in (\ref{eq:transmission_attempts}),
along with a DoS signal satisfying Assumptions 1 and 2. 	
If (\ref{co-located controller bound of DoS})
holds true, then the sequence of successful transmissions
satisfies $z_0 \leq Q$ and 
$z_{m+1}-z_{m} \leq Q + \Delta$ for all $m \in \mathbb N_0$, 
where
\begin{eqnarray}
Q=(\kappa + \eta \Delta) \left(1-\frac{1}{T} - \frac{\Delta}{\tau _D} \right)^{-1}
\end{eqnarray}
\qedp
\end{lemma}

\begin{figure}[t]
\begin{center}
\includegraphics[width=0.45 \textwidth]{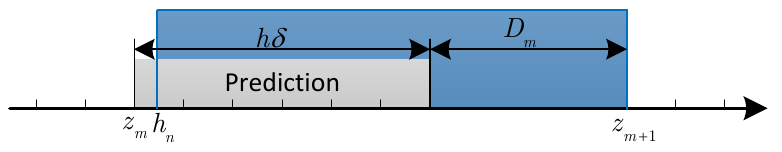} \\
\linespread{1}
\caption{\small Pictorial representation of (\ref{control input}) over
a generic interval $[z_m,z_{m+1}[$ where $m \in \mathbb N_0$. The
grey area represents the prediction horizon $h\delta$ during which the actuator utilizes the samples in the 
buffer, while the blue area exceeding $z_m+h\delta$ represents the period, say $|D_m|$, 
over which communication is not possible and there is no further control update.} 
\end{center}
\end{figure}

\subsection{Closed-loop behavior within the prediction horizon}
Lemma 1 guarantees that the time elapsing between successful transmissions
is always bounded under (\ref{co-located controller bound of DoS}). In particular, $z_0$ is finite. 
This makes it possible to focus the attention on the closed-loop behavior from $z_0$ onwards. 
Consider a generic interval $I_{p}(z_m)$ and let
\begin{eqnarray} \label{eq:error_dt}
\phi(t):= \alpha_p(z_m) -x(t), \quad t \in I_{p}(z_m)
\end{eqnarray} 

Notice that $I_{p}(z_m)$ might be empty, which happens 
whenever $z_{m+1} \leq z_m+p\delta$. Thus, in the sequel, we implicitly
consider only the intervals $I_{p}(z_m)$ which are nonempty. 
Exploiting (\ref{eq:error_dt}), we can rewrite the process dynamics as
\begin{eqnarray} \label{eq:system_equiv_form_dt}
\dot x(t)= \Phi x(t)+BK\phi(t)+d(t)
\end{eqnarray}

where $\Phi=A+BK$. Given any symmetric positive definite matrix 
$M$, let $P$ be the unique solution of 
the Lyapunov equation
$\Phi' P+P\Phi +M=0$. Let $V(x)=x' P x$.
Its derivative along the solutions to (\ref{eq:system_equiv_form_dt}), 
satisfies
{\setlength\arraycolsep{2pt} 
\begin{eqnarray} \label{eq:Lyap_dt}
\dot V(x(t)) \leq &-& \gamma _1 \|x(t)\|^2 + \gamma _2 \|x(t)\|\|\phi(t)\|  \nonumber \\
&+& \gamma _3 \|x(t)\|\|d(t)\|
\end{eqnarray}}%

where $\gamma _1$ is the smallest eigenvalue of $M$, $\gamma _2 :=\|2PBK\|$ and $\gamma _3 :=\|2P\|$. 
One sees that closed-loop stability depends on $\phi$, which, in turn, depends on the actuator 
sampling period $\delta$. In general, it is not possible 
to get a dissipation inequality in (\ref{eq:Lyap_dt}) for an arbitrary $\delta$.
Nonetheless, this is possible provided that $\delta$ is suitably chosen.

\begin{lemma}
Consider a dynamical system as in (\ref{system}) under the control action (\ref{control input}). 
Let the actuator sampling period $\delta$ be chosen as in Theorem 1. 
Then, there exists a positive constant 
$\rho$ such that
{\setlength\arraycolsep{2pt} 
\begin{eqnarray} \label{eq:error in buffer control}
\|\phi(t)\| \, \leq \, \sigma \|x(t)\| + \rho \left\| w_t \right\|_\infty
\end{eqnarray}}%

for all $t\in [z_m, \min\{z_m+h\delta,z_{m+1}\}[$, where $w=[d'\,\,n']'$.
\end{lemma}

\emph{Proof}. See Appendix A. \qedp

In view of Lemma 2, it is immediate to verify that if actuator sampling period is properly chosen 
then closed-loop stability is guaranteed since (\ref{eq:error in buffer control}) induces an 
ISS-type of property to the Lyapunov function. In fact, (\ref{eq:error in buffer control}) yields
{\setlength\arraycolsep{2pt} 
\begin{eqnarray}
\dot V(x(t)) \, &\le& \, - \, (\gamma _1 - \sigma \gamma _2) \|x(t)\|^2 \nonumber \\
&&\,  + \, (\gamma _2 \rho  + \gamma _3 ) \|x(t)\| \|w_t\|_\infty  \nonumber \\
&=&\, -\gamma_4 \|x(t)\|^2 + \gamma_5 \|x(t)\| \|w_t\|_\infty 
\end{eqnarray}}%

where $\gamma_4:=\gamma_1-\sigma \gamma_2$ 
and $\gamma_5:=\gamma _2 \rho  + \gamma _3$.

Observe now that for any positive real $\varepsilon$, the Young's inequality
\citep{Young} yields 
\begin{eqnarray} \label{eq:binomial}
2 \|x(t)\|  \|w_t\|_\infty \, \leq  \, \frac{1}{\varepsilon} \|x(t)\|^2 + \varepsilon \|w_t\|^2_\infty 
\end{eqnarray}

Using this inequality with $\varepsilon:=\gamma_5/\gamma_4$, 
we obtain
{\setlength\arraycolsep{2pt}
	\begin{eqnarray}
	\dot V(x(t)) &\leq& -\omega _1 V(x(t)) + \gamma_6 \|w_t\|^2_\infty   
	\end{eqnarray}}%
	
where $\omega _1:= \gamma _4/(2 \alpha_2)$ and 
$\gamma_6:= \gamma_5 ^2/(2 \gamma _4)$, 
where $\alpha_2$ denotes the largest eigenvalue of $P$.

Accordingly, 
{\setlength\arraycolsep{0pt} 
\begin{eqnarray} {\label{Lyapunov in buffered control}}
V(x(t)) \, &\le& \, e^{-\omega_1 (t-z_m)} V(x(z_m)) + \zeta_1  \| w_t\|^2_\infty
\end{eqnarray}}%

for all $t\in [z_m, \min\{z_m+h\delta,z_{m+1}\}[$, where $\zeta_1:=\gamma_6/\omega_1$.

\subsection{Closed-loop behavior outside the prediction horizon}

Lemma 1 guarantees that the time elapsing between successful transmissions
is always bounded under (\ref{co-located controller bound of DoS}). Specifically,
it ensures that $z_{m+1}-z_m\leq Q+\Delta$ for all $m\in \mathbb N_0$,
where $Q$ is a DoS-dependent constant. 
Consider now a transmission instant $z_m$. At this time, the buffer is full.
If $h \delta \geq \Delta$ and there is no DoS at $z_m+\Delta$ then the buffer is filled again
at $z_m+\Delta$ and the analysis of Section 4.2 applies. This scenario can be viewed as a co-located one. 
Thus, there are two critical cases: 
(i) when $h \delta < \Delta$ meaning that the prediction horizon does not cover one transmission period;
and (ii) when $\Delta \leq h \delta < \Delta + Q$, which means that the network is subject to
DoS periods that exceed the prediction horizon (\emph{cf.} Figure 3). We now discuss this scenario in detail.

Let $D_m:=[z_m+h\delta,z_{m+1}[$ with $D_m \neq \emptyset$.
Recall that the prediction error satisfies
\begin{eqnarray} \label{eq:prediction_out_buffer}
\phi(t) = \alpha_{h-1}(z_m)-x(t)
\end{eqnarray}  

for all $I_{h-1}(z_m)$. By hypothesis, $D_m \neq \emptyset$ which means that 
$z_{m+1}>z_m+h\delta$. This implies that (\ref{eq:prediction_out_buffer})
is valid for $t = z_m+h\delta$. Hence, the triangular inequality yields
\begin{eqnarray} {\label{error in DoS}}
\|\phi(t)\| &\le& \|\alpha_{h-1}(z_m)\| + \|x(t) \| \nonumber \\
&\le&  \|x(z_m+h\delta)\| + \|\phi(z_m+h\delta)\| +\| x(t) \|  
\end{eqnarray}

for all $t \in D_m$. We first look at $\phi (z_m+h\delta)$. Note that $\phi$ is continuous 
on $I_{h-1}(z_m)$. This, along with Lemma 2,
implies that 
\begin{eqnarray}
\|\phi(z_m+h\delta)\| \leq \sigma \|x(z_m+h\delta)\| + 
\rho \|w_{z_m+h\delta}\|_{\infty}
\end{eqnarray}

Combining this inequality with (\ref{error in DoS}), we get
\begin{eqnarray} {\label{error in DoS for Lyapunov}}
\|\phi(t)\| &\le&  (1+\sigma)\|x(z_m+h\delta)\| + \|x(t)\| + \rho \|w_t\|_\infty \nonumber \\
\end{eqnarray}

for all $t \in D_m$. 
Then, combining (\ref{error in DoS for Lyapunov}) and (\ref{eq:Lyap_dt}) yields
\begin{eqnarray} 
\dot V(x(t)) \, &\le& \,  (\gamma_2-\gamma_1) \|x(t)\|^2  \nonumber \\
			 && \,+ \, \gamma_2(1+\sigma)\|x(t)\|\|x(z_m+h\delta)\| \nonumber \\
			 &&\, + \, (\gamma_3+ \rho \gamma_2) \|x(t)\|\|w_t\|_\infty  
\end{eqnarray}

By applying again the Young's inequality 
with $\varepsilon:=\frac{\gamma_3+ \rho \gamma_2}{\gamma_1 - \sigma \gamma_2}$, we obtain
\begin{eqnarray}
\dot V(x(t)) &\le& \gamma_2 \|x(t)\|^2  \nonumber \\
&&+ \gamma_2(1+\sigma)\|x(t)\|\|x(z_m+h\delta)\| + \gamma_7\|w_t\|^2_\infty \nonumber \\
\end{eqnarray}
where $\gamma_7: = \frac{(\gamma_3+ \rho \gamma_2)^2}{2(\gamma_1-\sigma\gamma_2)}$. 

Hence, for every $t\in D_m$ such that $\|x(t)\|\ge \|x(z_m+h\delta)\|$ we have
\begin{eqnarray} {\label{Lyapunov in DoS situation 1}}
\dot V(x(t)) \, \leq \, \omega_2 V(x(t)) + \gamma_7 \|w_t\|^2_ \infty
\end{eqnarray} 

where $\omega_2:= \gamma_2(2+\sigma)/\alpha_1$. 
On the other hand, for every $t\in D_m$ such that  $\|x(t)\| \le \|x(z_m+h\delta)\|$ we have
\begin{eqnarray} {\label{Lyapunov in DoS situation 2}}
\dot V(x(t)) \, \leq \, \omega_2 V(x(z_m+h\delta)) + \gamma_7 \|w_t\|^2_ \infty
\end{eqnarray}

From these expressions, we 
conclude that during each interval $D_m$ exceeding the prediction horizon 
the Lyapunov function satisfies
\begin{eqnarray} {\label{Lyapunov function in DoS}}
V(x(t)) \, &\le& \, e^{\omega_2(t-z_m-h\delta)} V(x(z_m+h\delta))  \nonumber \\
   && \, + \zeta_2  e^{\omega_2(t-z_m-h\delta)} \|w_t\|^2_ \infty
\end{eqnarray}

where $\zeta_2:= \gamma_7/\omega_2$.

\subsection{Stability analysis}
The foregoing analysis indicates that the overall behevior of the system 
can be regarded as the one of a switched system that behaves stably 
over the intervals $[z_m,z_m+h\delta[$ and unstably over the intervals $D_m$.
We have the following result.

\begin{theorem}
Consider a dynamical system as in (\ref{system}) under the control action (\ref{control input}). 
Given any positive definite symmetric matrix $Q$,
let $P$ denote the solution of the Lyapunov equation 
$\Phi' P + P \, \Phi + Q = 0$ with $\Phi=A+BK$ stable.
Let the actuator sampling period $\delta$ be as in Theorem 1. Consider any DoS pattern
satisfying (\ref{co-located controller bound of DoS}).
Then, the closed-loop system is stable if the prediction horizon $h\delta$ satisfies
\begin{eqnarray} {\label{predictioni horizon}}
h \delta > \frac{\omega_2}{\omega_1+\omega_2} (Q+\Delta)
\end{eqnarray}

where $Q$ is as in Lemma 1, $\omega_1= \frac{\gamma_1-\sigma\gamma_2}{2\alpha_2}$ 
and $\omega_2= \frac{\gamma_2(2+\sigma)}{\alpha_1}$ where $\gamma_1$, $\gamma_2$ and $\sigma$
are as in Theorem 1 and where $\alpha_1$ and $\alpha_2$ are the smallest and largest 
eigenvalues of $P$, respectively.
\end{theorem}

\emph{Proof.}
Recall that $z_0$ exists finite in view of Lemma 1. Hence, we can restrict 
ourselves to study the closed-loop behavior from $z_0$ onwards. 
We also assume without loss of generality that $h\delta<Q+\Delta$
otherwise the result follows immediately from the analysis of Section 4.2.
We show that 
\begin{eqnarray} \label{eq:0}
V(x(z_m)) \leq  \lambda e^{-\beta (z_m-z_0)} V(x(z_0)) + 2 \zeta(z_m) \sum_{k=0}^{m} L^k \nonumber \\
\end{eqnarray}
where $\beta$ is defines as
\begin{eqnarray} \label{eq:beta}
\beta := \frac{\omega_1h\delta - \omega_2(Q+\Delta-h\delta)}{Q+\Delta}
\end{eqnarray}

and where
\begin{eqnarray}
&& \lambda := e^{(\omega_1+\omega_2)Q} \\ 
&& L:= e^{-\beta \Delta} \\
&& \zeta(t) : = 2 \max \{\zeta_1,\zeta_2\} \,
e^{\omega_2(Q+\Delta-h\delta)} \|w_{t}\|^2_ \infty
\end{eqnarray}

Note that $\beta>0$ in view of (\ref{predictioni horizon}). 

We prove this claim through an induction argument.
The claim trivially holds for $m=0$ since $\lambda \geq 1$.
Assume next that the claim is true up to $z_m$ with $m \in \mathbb N_1$. 
First recall that 
(\ref{Lyapunov in buffered control}) implies 
{\setlength\arraycolsep{2pt} 
\begin{eqnarray} \label{eq:1}
&& V(x(t)) \, \leq \,  e^{-\omega_1 (t-z_m)} V(x(z_m)) + \zeta(t) \nonumber \\ \nonumber \\
&& \quad \leq \,  \lambda e^{-\beta(t-z_0)} V(x(z_0)) 
+2  e^{-\beta(t-z_m)} \zeta(t) \sum_{k=0}^{m} L^k +\zeta(t) \nonumber \\ 
\end{eqnarray}}%
for all $t\in [z_m, \min\{z_m+h\delta,z_{m+1}\}[$, where the second inequality follows from $\omega_1 > \beta$.
Then, we have two cases. Assume $D_m=\emptyset$, which implies that $z_{m+1} \leq z_m+h\delta$.
Note that $z_{m+1}-z_m \geq \Delta$ by construction.
Hence, at time $z_{m+1}$ we have $e^{-\beta(z_{m+1}-z_m)} \leq e^{-\beta \Delta} =L$.
This shows that (\ref{eq:0}) holds true at $z_{m+1}$. 
Assume next $D_m \neq \emptyset$.
In view of (\ref{Lyapunov in buffered control}) and since $V$ is continuous, we get
{\setlength\arraycolsep{2pt} 
\begin{eqnarray}
V(x(z_m+h\delta)) &\leq& e^{-\omega_1 h\delta} V(x(z_m)) + \zeta_1 \|w_{z_m+h\delta}\|^2_\infty \nonumber \\
\end{eqnarray}}%
Combining this inequality with (\ref{Lyapunov function in DoS}) yields
{\setlength\arraycolsep{2pt} 
\begin{eqnarray} \label{eq:2}
V(x(t)) \, &\leq& \, e^{\omega_2 (t-z_m-h\delta)} V(x(z_m+h\delta)) + \zeta(t)  \nonumber \\ 
&\leq& \, e^{\omega_2 (t-z_m-h\delta)} 
[ e^{-\omega_1 h\delta} V(x(z_m)) + \zeta(t) ] + \zeta(t) \nonumber \\ 
&\leq& \,e^{-\beta ( Q + \Delta) } V(x(z_m)) + 2 \zeta(t)  \nonumber \\ 
&\leq& \, \lambda e^{-\beta (t-z_0)} V(x(z_0)) + 2 \zeta(t) \sum_{k=0}^{m+1} L^k
\end{eqnarray}}%

for all $t \in D_m$, where the third inequality follows from the fact that $t \leq z_m + Q + \Delta$
and the definition of $\beta$, while the last inequality follows from the fact that $e^{-\beta ( Q + \Delta) }\leq L$
and the fact that $Q + \Delta+z_m \geq z_{m+1}\geq t$ for all $t \in D_m$.
Thus, (\ref{eq:0}) holds true at $z_{m+1}$.

Using this property, the proof can be easily finalized. In fact, 
(\ref{eq:1}) and (\ref{eq:2}) yield
{\setlength\arraycolsep{2pt} 
\begin{eqnarray}
V(t) &\leq& \lambda e^{-\beta (t-z_0)} V(x(z_0)) + 2 \zeta(t) \sum_{k=0}^{m+1} L^k + \zeta(t) 
\end{eqnarray}}%
for all $[z_m,z_{m+1}[$, and the sum term is bounded for any $m$ since $L<1$. 
This concludes the proof. \qedp

\begin{remark}
One sees from (\ref{predictioni horizon}) that in order to get stability it is not necessary that 
the prediction horizon covers the maximum period of DoS. In fact, also large DoS periods 
can be tolerated. In particular, one sees that robustness increases with: (i) small values of $\omega_2$,
which corresponds to mild open-loop unstable dynamics; and (ii) large values of $\omega_1$,
which corresponds to a large decay rate of the DoS-free closed-loop dynamics. The latter 
should be therefore taken into account when designing the state-feedback matrix $K$.
\qedp
\end{remark}

\begin{remark}
Taking into account the expression of $Q$ from Lemma 1, it is immediate to see that 
(\ref{predictioni horizon}) can be equivalently rewritten as (\ref{eq:stability_condition_equivalent}).
This shows that the considered architecture approaches the ideal bound (\ref{co-located controller bound of DoS})
as $h \rightarrow \infty$. Inequality (\ref{eq:stability_condition_equivalent}) does also quantify the ``gap"
between remote and co-located architectures for given any finite $h$.
\qedp
\end{remark}

\begin{remark}
In this paper, we have assumed that the time needed to compute $h$ control values is zero.
This assumption can be restrictive if $h$ is large. As discussed in \cite{7526102},
Lemma 1 follows from the fact that every time interval of duration $Q + \Delta$ contains at least 
a DoS-free interval of duration $\Delta$. 
This result can be generalized. 
In fact, one can show that if we replace
(\ref{co-located controller bound of DoS}) with 
\begin{eqnarray}
\frac{1}{T} + \frac{\mu \Delta}{\tau _D} < 1, \quad \mu \in \mathbb N_1
\end{eqnarray}

then every time interval of duration 
\begin{eqnarray}
Q_\mu := (\kappa + \eta \mu \Delta) \left(1-\frac{1}{T} - \frac{\mu \Delta}{\tau _D} \right)^{-1}
\end{eqnarray}
contains at least 
a DoS-free interval of duration $\mu \Delta$. 
This property is very useful to account for non-zero computation times $T_c$.
In fact, this property makes it possible to regard $T_c$ as an extended DoS interval.
The only modification is that the sequence of control values should start 
from $\alpha_{\left \lceil{T_c/\Delta}\right \rceil \Delta}$.
\qedp
\end{remark}

\section{A Numerical Example}

The numerical example is taken from \citep{forni}. The system to be controlled
is open-loop unstable and is characterized by the matrices 
\begin{eqnarray} \label{eq:system_example}
	A=\left[ \begin{array}{cc} \, 1 \, & \, 1 \, \\ 0 \, &  \, 1 \end{array} \right], \quad 
	B= \left[ \begin{array}{cc} \, 1 \, & \, 0 \, \\ 0 \, &  \, 1 \end{array} \right]
\end{eqnarray} 

The state-feedback matrix is given by
\begin{eqnarray} \label{eq:controller_example}
	K = \left[ \begin{array}{cc} -2.1961 & -0.7545 \\ -0.7545 & -2.7146 \end{array} \right]
\end{eqnarray} 

The control system parameters are $\gamma_1=1$, $\gamma_2=2.1080$, 
$\alpha_1=0.2779$, $\alpha_2=0.4497$, $\|\Phi\|=1.9021$, $\omega_1=0.5025$, $\omega_2=15.1709$ and $\mu_A=1.5$. 
Disturbance $d$ and noise $n$ 
are random signals with uniform distribution in $[-0.01,0.01]$. 

The network transmission rate is given by $\Delta=0.1$s. 
As for the actuator sampling period, Theorem 1 yields $\delta < 0.1508$.
We select $\delta=0.1$s in order to synchronize the 
controller sampling rate with $\Delta$. 

Figure 4 shows simulation results comparing the 
co-located architecture in \citep{7526102} with the one proposed in this paper.
We consider a sustained DoS attack with variable period and duty cycle, generated randomly.
Over a simulation horizon of $50$s, the DoS 
signal yields $|\Xi(0,50)|=34.65$s and $n(0,50)=39$. 
This corresponds to values (averaged over $50$s) 
of $\eta \le 3.1$, $\kappa \le 0.8442$, $\tau_D\approx1.2821$ and $T\approx1.4430$, and $\sim70\%$ of transmission failures.
Moreover,
\begin{eqnarray} {\label{simulatioin bound}}
\frac{\Delta}{\tau_D} + \frac{1}{T} \approx 0.7710
\end{eqnarray}

For the co-located architecture, the requirement for closed-loop stability is clearly satisfied.

Consider next the remote architecture. In agreement with (\ref{predictioni horizon}), to get stability 
one needs a prediction horizon $h\delta \geq 4.9153$, which corresponds to $h \geq 50$.
One sees from Figure 4 that without buffering, namely when $h=1$, the closed-loop system is unstable.
On the other hand, $h=5$ is sufficient to obtain a satisfactory closed-loop response. 
The results indicate that in some cases, like the one discussed here, small values of $h$
can approximate well the co-located architecture without actually requiring large $h$ or co-location.
In this sense, the proposed control architecture provides an effective solution to
trade-off ease of implementation and robustness against DoS. 
 
The conservativeness of the theoretical bound on $h$ is somewhat 
implicit to the ``worst-case" type of analysis pursued here. 

\begin{figure}[tb]
	\begin{center}
		\psfrag{x1}{{\tiny $x_1$}}
		\psfrag{x2}{{\tiny $x_2$}}
		\psfrag{DoS}{{\tiny DoS}}
		\includegraphics[width=0.45 \textwidth]{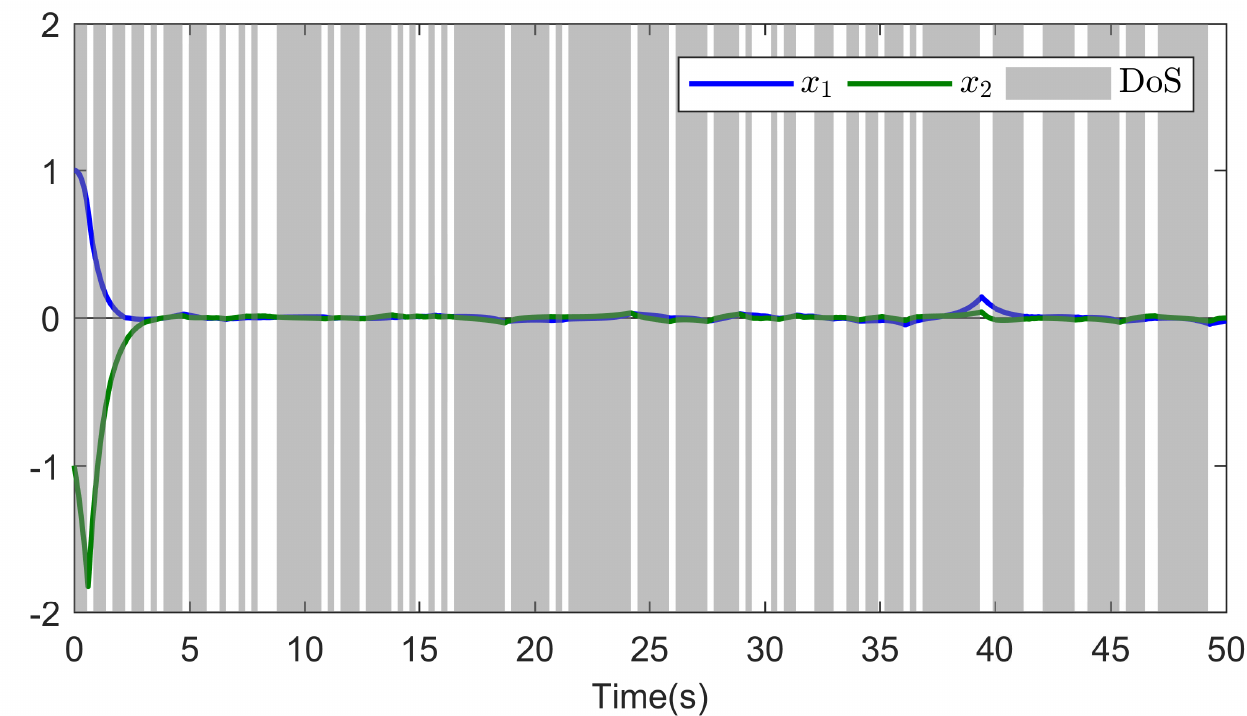} \\
		\includegraphics[width=0.45 \textwidth]{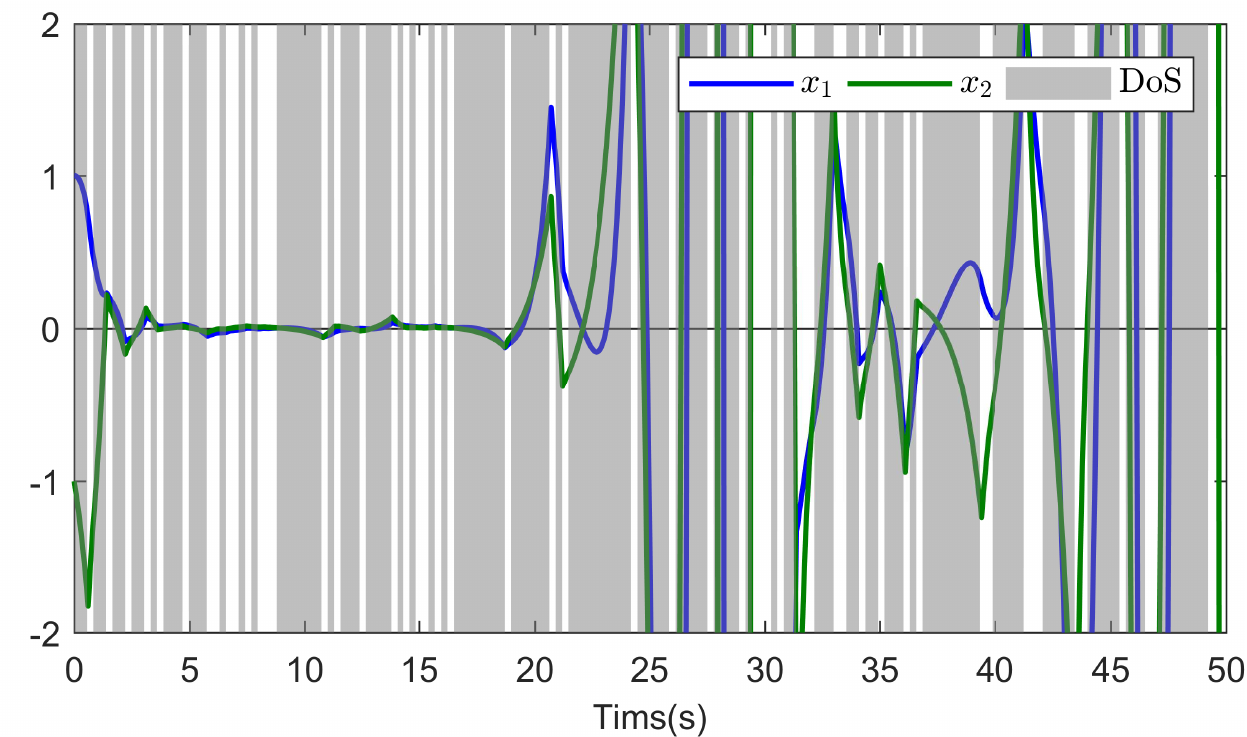} \\
		\includegraphics[width=0.45 \textwidth]{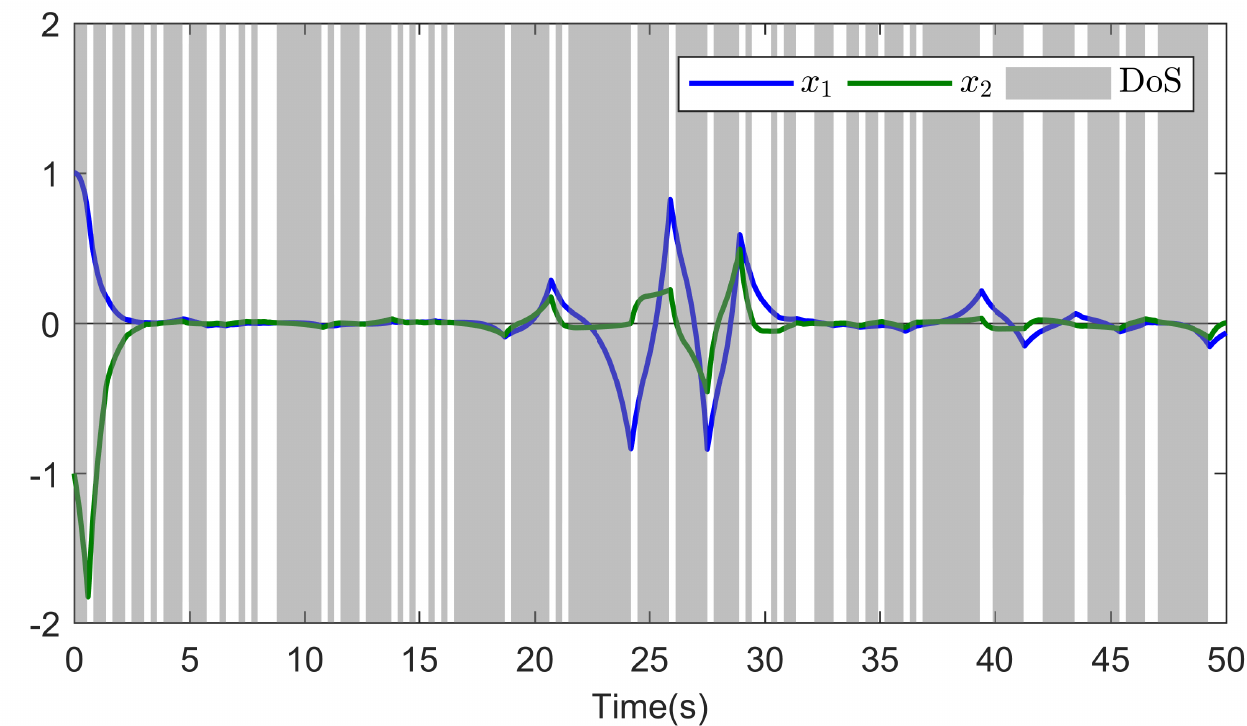} \\
		\linespread{1}\caption{\small Simulation results for the example 
			in case disturbance and noise are random signals with uniform distribution in $[-0.01, 0.01]$.
			Top: Co-located architecture; Middle: Remote architecture without buffering; 
			Bottom: Remote architecture with buffering ($h=5$).} \label{fig:2}
	\end{center}
\end{figure}

\section{Concluding remarks}
In this paper, we have investigated networked systems in the presence 
of Denial-of-Service attacks, comparing co-located and remote control 
architectures. While the former achieves the highest level of robustness
for a very large class of DoS signals, it need not be always feasible from a practical point of view.
We introduced a control architecture that approximates co-location while enabling remote implementation.
This architecture relies on ``packet-based" control: the control unit transmits a sequence of controls 
containing process input predictions. Whenever communication is allowed, 
the controls are stored in a buffer and used during the DoS periods. 
We studied closed-loop stability, quantifying the robustness ``gap"
between this architecture and the co-located one. As it emerges from both the analysis and the numerical simulations,
the proposed architecture provides an effective solution for
trading-off ease of implementation and robustness against DoS. 

The present results can be extended in many directions. The case
of partial state measurements represents an interesting research venue, where the results in \citep{Feng201742} may prove relevant in this regard.
Another interesting research line pertains the study of remote control architecture
when sensor and actuator channels are independent. In this case, it is interesting to
study closed-loop stability in connection with asynchronous DoS.

\appendix

\section{Proof of Lemma 2}

Consider any interval $[z_m,\min\{z_m+h\delta,z_{m+1}\}[$, 
$m \in \mathbb N_0$. The proof is divided into two steps. 
First, we provide an upper bound on the error dynamics $\phi$ at the sampling times $z_m+p\delta$.
Second, we provide an upper bound on the error dynamics $\phi$
between inter-samplings. This provides 
an upper bound on $\phi$ over the whole interval $[z_m,\min\{z_m+h\delta,z_{m+1}\}[$.

We start by deriving an upper bound on $\|\phi(z_m+p\delta)\|$.
It is simple to verify that
the dynamics of the variable $\alpha$ 
related to the predictor equation satisfies 
{\setlength\arraycolsep{2pt} 
\begin{eqnarray}
\alpha_p(z_m) \, = \, A_\delta^p  \, \alpha_0(z_m)  
+ \sum_{k=0}^{p-1}  A_\delta^{p-k-1}  B_\delta K\, \alpha_k(z_m) \nonumber \\
\end{eqnarray}}%

for all $p \in \{0,1,\ldots,h-1\}$ such that $z_m+p\delta < z_{m+1}$. Moreover, 
the process dynamics satisfies
{\setlength\arraycolsep{2pt} 
\begin{eqnarray}
x(t) \, &=& \, e^{A (t-z_m) } x(z_m) + \int_{z_m}^t   e^{A (t-\tau) } B u(\tau) d \tau
 \nonumber \\ 
&& +  \, \int_{z_m}^t   e^{A (t-\tau) } d(\tau) d \tau
\end{eqnarray}}%

for all $t \in \mathbb R_{\geq z_m}$. Combining these two expressions, we get
{\setlength\arraycolsep{1pt} 
\begin{eqnarray} 
\phi(z_m+p\delta ) &=& \alpha_p(z_m) - x(z_m+p\delta ) \nonumber \\
&=& e^{A p \delta}n(z_m)-\int_{z_m}^{z_m+p\delta}
e^{A(z_m+p\delta  -\tau)}d(\tau)d\tau  \nonumber \\
\end{eqnarray}}%
for all $p \in \{0,1,\ldots,h-1\}$ such that $z_m+p\delta < z_{m+1}$.
Here, we exploited 
the relation $A_\delta^p=e^{A p \delta }$ and the fact that {\setlength\arraycolsep{2pt} 
\arraycolsep=1.4pt\def\arraystretch{2}
\begin{eqnarray}
&& \int_{z_m}^{z_m+p\delta }   e^{A (z_m+p\delta -\tau) } 
B u(\tau) d \tau  \nonumber \\
&& \quad  = \sum_{k=0}^{p-1} \left[ \int_{z_m+k\delta}^{z_m+(k+1) \delta} 
e^{A(z_m+p\delta -\tau)} B d\tau  
\right] K\alpha_k(z_m) \nonumber \\
&& \quad  = \sum_{k=0}^{p-1}  e^{A \delta (p - k  -1)}
\left[ \int_{0}^{\delta} 
e^{A s} B d s 
\right] K \alpha_k(z_m) 
\nonumber \\
&& \quad  = \sum_{k=0}^{p-1}  A_\delta^{p - k  -1}
B_\delta K \alpha_k(z_m) 
\end{eqnarray}}%
where the second equality is obtained using the change of variable
$s = z_m + (k+1)\delta -\tau$.
We then have 
{\setlength\arraycolsep{2pt} 
\begin{eqnarray} 
\|\phi(z_m+p\delta )\| &\leq& \rho_1 \|w_t\|_\infty
\end{eqnarray}}

where $\rho_1:=(1+\frac{1}{\mu_A})e^{\mu_A(h-1)\delta}$. 

We can now provide an upper bound on the prediction error $\phi$
between inter-sampling instants. To this end, observe that 
the dynamics of $\phi$ satisfies
{\setlength\arraycolsep{2pt} 
	\begin{eqnarray}  \label{eq:error_dynamics_dt}
		\arraycolsep=1pt\def\arraystretch{1.5}
	\dot \phi(t) \, &=& \,  - \dot x(t) \nonumber \\
	&=& \,  A \phi(t) - \Phi \alpha_p(z_m) - d(t), \quad t \in I_{p}(z_m) 
\end{eqnarray}}%

where $\phi(z_m) = n(z_m)$. Let now
{\setlength\arraycolsep{2pt} 
\begin{eqnarray} 
f(t-z_m-p\delta ):=\int_{z_m+p\delta}^{t} e^{\mu_A(t-\tau)}  d\tau
\end{eqnarray}}%

where $t \in \mathbb R_{z_m+p\delta}$.
Then, the solution of (\ref{eq:error_dynamics_dt})
satisfies 
{\setlength\arraycolsep{3pt} 
\begin{eqnarray} 
\arraycolsep=2.4pt\def\arraystretch{3}
\|\phi(t)\| \, & \leq & \, e^{\mu_A(t-z_m-p\delta )}  \|\phi(z_m+p\delta )\|  \nonumber \\
&&  \, + \, \kappa_1 f(t-z_m-p\delta ) ( \|d_t\|_\infty +  \|\alpha_p(z_m)\| ) \nonumber \\  
&\leq& \, \rho_1 \rho_2  \|w_t\|_\infty + \, \kappa_1 f(t-z_m-p\delta ) \|d_t\|_\infty  \nonumber \\  
&& \, + \, \kappa_1 f(t-z_m-p\delta ) ( \|\phi(t)\| + \|x(t)\| ) 
\end{eqnarray}}%

for all $t \in I_p(z_m)$, where we defined $\rho_2 := \max \{ e^{\mu_A \delta},1 \}$
and $\kappa_1:=\max\{\|\Phi\|,1\}$.

Observe that $f(0)=0$ and that $f(t-z_m-p \delta)$ is monotonically increasing with $t$. 
Thus, any positive real $\delta$ such that
\begin{eqnarray} \label{eq:Delta_bar}
f(\delta) \,  \leq \, \frac{1}{\kappa_1} \frac{\sigma}{(1+\sigma)} 
\end{eqnarray}

ensures (\ref{eq:error in buffer control}) 
with $\rho : = \sigma + \rho_1 \rho_2 (1+\sigma)$.

The explicit expression for $\delta$ follows from Theorem 1. \qedp

\bibliography{ref}

\end{document}